# Interplay between moment-dependent and field-driven unidirectional magnetoresistance in CoFeB/InSb/CdTe heterostructures

Jiuming Liu[1], Liyang Liao[2], Bin Rong[1], Yuyang Wu[3], Yu Zhang[4], Hanzhi Ruan[1], Zhenghang Zhi[1,5,6], Puyang Huang[1], Shan Yao[1], Xinyu Cai[7], Chenjia Tang[7,8], Qi Yao[7,8], Lu Sun[1], Yumeng Yang[1], Guoqiang Yu[4], Renchao Che[3*], and Xufeng Kou[1,8*]

[1]School of Information Science and Technology, ShanghaiTech University, Shanghai, 201210, China

[2]Institute for Solid State Physics, University of Tokyo, Kashiwa 277-8581, Japan

[3]Laboratory of Advanced Materials, Shanghai Key Lab of Molecular Catalysis and Innovative Materials, Fudan University, Shanghai 200438, China

[4]Beijing National Laboratory for Condensed Matter Physics Institute of Physics Chinese Academy of Sciences Beijing 100190, China

[5]Shanghai Institute of Microsystem and Information Technology, Chinese Academy of Sciences, Shanghai 200050, China

[6]University of Chinese Academy of Science, Beijing 101408, China

[7]School of Physical Science and Technology, ShanghaiTech University, Shanghai, 201210, China

[8]ShanghaiTech Laboratory for Topological Physics, ShanghaiTech University, Shanghai 201210, China

*Correspondence to: rchche@fudan.edu.cn; kouxf@shanghaitech.edu.cn

**Magnetoresistance effects are crucial for understanding the charge/spin transport as well as propelling the advancement of spintronic applications. Here we report the coexistence of magnetic moment-dependent (MD) and magnetic field-driven (FD) unidirectional magnetoresistance (UMR) effects in CoFeB/InSb/CdTe heterostructures. The strong spin-orbital coupling of InSb and the matched impedance at the CoFeB/InSb interface warrant a distinct MD-UMR effect at room temperature, while the interaction between the in-plane magnetic field and the Rashba effect at the InSb/CdTe interface induces the marked FD-UMR signal that dominates the high-field region. Moreover, owning to the different spin transport mechanisms, these two types of nonreciprocal charge transport show opposite polarities with respect to the magnetic field direction, which further enable an effective phase modulation of the angular-dependent magnetoresistance. Besides, the demonstrations of both the tunable UMR response and two-terminal spin-orbit torque-driven magnetization switching validate our CoFeB/InSb/CdTe system as a suitable integrated building block for multifunctional spintronic device design.**

Magnetoresistance (MR), which describes the variation of the electrical resistance in response to the applied magnetic field, plays an essential role in spin-related physics and applications. For example, the giant magnetoresistance (GMR)[1,2] and tunneling magnetoresistance (TMR) effects have revolutionized the non-volatile memory industry[3-5]. In the meantime, the anisotropic magnetoresistance (AMR)[6, 7] has empowered remarkable advancements in sensor technology[8,9]. Recently, it is suggested that the interaction between the electron spin and magnetism can give rise to a new type of MR effect called unidirectional magnetoresistance (UMR)[10,11]. Governed by the spin-induced asymmetry argument, such UMR effect is odd with one-fold symmetry under the reversal of either the charge current *j* or magnetic-related parameters (magnetic moment *M* or magnetic field *B*)[12,13]. Depending on the origin of UMR, it is found that the transport scattering of polarized spin that comes from either the bulk-type spin Hall effect or Rashba effect in the large spin-orbit coupling (SOC) channel would be affected by the magnetization of the adjacent magnetic layer, hence leading to the magnetic moment-dependent unidirectional magnetoresistance (MD-UMR) effect. Accordingly, the discovered unidirectional spin hall magnetoresistance[10], unidirectional Rashba-Edelstein magnetoresistance[14], and anomalous unidirectional magnetoresistance[15, 16] in various magnetic heterostructures all belong to this category, and their corresponding 2$^{nd}$-harmonic resistances $R^{2\omega}$ are proportional to $j \times M$. Benefiting from both the two-terminal write/read mechanism and high spin-orbit torque (SOT) switching efficiency, the MD-UMR-based devices show great potentials in non-volatile application[17, 18]. Following the same spin interaction scenario, when the time reversal symmetry is broken by a magnetic field *B*, the spin-momentum locking mechanism originated from the Rashba effect or topological surface states can also lead to a non-reciprocal charge transport in non-centrosymmetric systems, and the resulting magnetic field-driven unidirectional magnetoresistance (FD-UMR) is characterized by $R^{2\omega} \sim j \times B$[11, 19, 20]. Apart from probing the mesoscopic spin texture of the host material[11], the FD-UMR effect can also add a 360° directional sensitivity to AMR-based sensors[10] and additional amplitude-detection capabilities to MD-UMR based sensors. In this regard, it would take advantages to integrate both UMR effects into one platform for constructing multi-functional spintronic devices.

To implement the aforementioned proposal, in this work, we report the use of the narrow bandgap semiconductor-based magnetic heterostructures, CoFeB/InSb/CdTe, to tailor the non-reciprocal electrical response through the interplay between the MD- and FD-UMR contributions.

In this multi-layer system, two spin-dependent transport channels are identified from the systematic magneto-transport measurements. In particular, the large intrinsic SOC of InSb[21] and its comparable electrical resistivity (~7.7 × $10^{-3}$ Ω·cm) to CoFeB (~1.4 × $10^{-4}$ Ω·cm) warrant an efficient spin accumulation at the CoFeB/InSb interface, which in turn gives rise to a pronounced MD-UMR effect with the amplitude of the $2^{nd}$-harmonic resistance 3~5 times higher than other magnetic heterostructures. Meanwhile, the giant Rashba-type SOC owning to the effective quantum confinement and band bending at the InSb/CdTe hetero-interface introduces a room-temperature FD-UMR signal which dominates in the high-field region. Moreover, the opposite polarities of these two UMR phenomena enable the transition of the angular-dependent $2^{nd}$-harmonic resistance curves where the peak and valley positions swap under different magnetic fields. In addition, we demonstrate the electrical write and read operations of the two-terminal SOT prototype device based on the CoFeB/InSb/CdTe framework.

Experimentally, high-quality InSb/CdTe heterostructures were firstly prepared by molecular beam epitaxy (MBE)[22]. A two-step growth strategy was adopted during the CdTe buffer layer growth, where an initial low-temperature growth mode helped to rapidly release the stress introduced by the mismatch between the CdTe and GaAs lattices, while the subsequent growth at a higher temperature promoted the adatom surface migration to ensure a smooth surface morphology. Accordingly, the matched lattice constants between InSb and CdTe resulted in synthesizing a single-crystalline and defect-free InSb layer with an atomically sharp interface that serves as the foundation for strong interfacial Rashba SOC[23]. Afterwards, the CoFeB(5nm) layer with an in-plane anisotropy (IMA) was deposited using magneton sputtering, followed by the capping of the Ta(1nm)/MgO(2nm) film stack to prevent the surface oxidation. After sample growth, μm-sized six-terminal Hall bar devices were fabricated using standard photolithography and ion-beam etching methods[23]. The etching process stopped at the insulating CdTe layer to ensure that the electron current only flew through the CoFeB/InSb channel, and the Ti/Au (20nm/280nm) electrodes were deposited using e-beam evaporation to form Ohmic contacts. Besides, considering that the MD-UMR measurements require the spin polarization to be parallel or anti-parallel with the magnetic moment, the longitudinal direction of the Hall-bar device was defined perpendicular to the easy axis of the CoFeB layer, as shown in Fig.1a. To characterize the magnetic properties of the CoFeB/InSb/CdTe system, MOKE microscopy was carried out at room temperature. By sweeping the in-plane magnetic field from –5mT to 5mT along the *y*-direction, Fig.1b displays the field-dependent MOKE intensity data of the Hall-bar device. The magnetic hysteresis loops exhibit a nearly square shape with the coercivity field of $H_C$ = 2mT, manifesting the dominant in-plane anisotropy in the CoFeB layer. Besides, the uniform red and blue patterns of the MOKE images (inset of Fig. 1b) throughout the device channel signify the fully magnetized CoFeB layer along the +*y* ($M_y \uparrow$) and −*y* ($M_y \downarrow$) directions under high magnetic fields, respectively.

The MD-UMR effect of the CoFeB(5nm)/InSb(15nm)/CdTe(400nm) device was investigated by angle-dependent magneto-transport measurements. By successively rotating the applied magnetic field of |***B***|= 50 mT in the *x-y* plane and an a.c. stimulated current of *I* = 1.5mA (*x*-direction, frequency = 17.8Hz), both the 1$^{st}$ and 2$^{nd}$ harmonic components of the longitudinal ($R_{xx}^\omega$, $R_{xx}^{2\omega}$) and Hall ($R_{xy}^\omega$, $R_{xy}^{2\omega}$) resistances were recorded, and Figs. 2a-b display relevant results at room temperature. Consistent with heavy metal-based magnetic heterostructures[10, 24], both $R_{xx}^\omega$ and $R_{xy}^\omega$ show typical AMR and planar hall effect (PHE) behaviors with a two-fold symmetry (*i.e.*, the periods of the AMR and PHE curves are both 180°). On the contrary, the $R_{xx}^{2\omega}$ and $R_{xy}^{2\omega}$ exhibit sinusoidal dependencies in reference to the rotation angle $\phi$ (*i.e.*, the angle between ***B*** and *I*) with the period of 360°: for instance, it is seen that $R_{xx}^{2\omega}$ reaches its minimum and maximum values at $\Phi$=90˚ and 270˚, respectively. In addition to the angular-dependent magnetoresistance results, Fig. 2c presents the complementary low-field $R_{xx}^{2\omega} - B$ curves with the $B \perp I$ configuration under three different current levels of $I$ = 0.5, 1.0, and 1.5mA. Specifically, $R_{xx}^{2\omega}$ reaches the low (high) state when magnetic moment $M$ is parallel (anti-parallel) to the spin polarization $\sigma$ (*i.e.*, following the spin Hall effect and interfacial Rashba scenario, the charge current generates the $+\sigma_y$ condition at the CoFeB/InSb interface (Fig. 2d), and our results are in agreement with the reported ferromagnet (FM)/normal metal (NM)[10] and FM/topological insulator(TI)[14] systems). Meanwhile, the amplitude of the two-state hysteresis loop $|\Delta R_{xx}^{2\omega}|$ is found to positively correlate with the applied current *I*, whereas the hysteresis windows remain almost constant regardless of the current amplitude variation.

Here, it is noticed that both the thermoelectric effect ($R_{xx}^{2\omega,\text{th}} \sim j \cdot (M \times \nabla T)$) and MD-UMR effect ($R_{xx}^{2\omega,\text{MD-UMR}} \sim j \times M$) would contribute to the overall 2$^{\text{nd}}$-harmonic signals[10,25,26]. Quantitatively, the angle-dependent $R_{xx}^{2\omega}$, $R_{xy}^{2\omega}$ curves can be expressed as:

$$R_{xx}^{2\omega} = R_{xx}^{2\omega,\text{FL}}\sin2\phi\cos\phi + (R_{xx}^{2\omega,\text{th}} + R_{xx}^{2\omega,\text{MD-UMR}})\sin\phi \quad (1)$$

$$R_{xy}^{2\omega} = R_{xy}^{2\omega,\text{FL}}(2\cos^3\phi - \cos\phi) + (R_{xy}^{2\omega,\text{th}} + R_{xy}^{2\omega,\text{DL}})\cos\phi \quad (2)$$

where $R_{xx}^{2\omega,\text{FL}}, R_{xx}^{2\omega,\text{th}}, R_{xy}^{2\omega,\text{FL}}, R_{xy}^{2\omega,\text{th}}, R_{xy}^{2\omega,\text{DL}}$ refer to the field-like SOT (FL-SOT), thermal effect components of $R_{xx}^{2\omega}$, and FL-SOT, thermal effect, and damping-like SOT (DL-SOT) components of $R_{xy}^{2\omega}$, respectively[10]. Based on Eq. (1), it is seen that $R_{xy}^{2\omega,FL}$ can be excluded from the $\sin\phi$-dependent component of the measured $R_{xy}^{2\omega}$ data (Supplementary Note1). To evaluate the thermal contribution, we subsequently recorded a set of $R_{xy}^{2\omega} - \phi$ data points under different in-plane magnetic field and current combinations. Figure 2e visualizes the three-dimensional $\Delta R_{xy}^{2\omega} = R_{xy,1}^{2\omega}|_{\phi=0°} - R_{xy,1}^{2\omega}|_{\phi=180°}$ mapping (where $R_{xy,1}^{2\omega} = R_{xy}^{2\omega,\text{th}} + R_{xy}^{2\omega,\text{DL}}$ in reference to Eq. (2)), in which all data points are around a slanted plane, implying the presence of the heating contribution (Supplementary Note2). Next, considering that $R_{xy}^{2\omega,\text{DL}}$ is inversely proportional to $1/(B + B_{\text{dem}} - B_{\text{ani}})$[10, 14] where $B_{\text{dem}}$ and $B_{\text{ani}}$ are the demagnetization field and the perpendicular anisotropic field of the CoFeB layer (*i.e.*, the quantification of $B_{\text{dem}} - B_{\text{ani}}$ is discussed in Supplementary Note3, we further plotted the $R_{xy}^{2\omega,\text{th}+\text{DL}}$ versus $1/(B + B_{\text{dem}} - B_{\text{ani}})$ (*i.e.*, which is obtained by subtracting the linear background of the measured $\Delta R_{xy}^{2\omega}$-$\boldsymbol{B}$ data of Fig. 2e), and the intercept of the linear fitting line across the *y*-axis corresponds to the thermal contribution $R_{xy}^{2\omega,\text{th}} = R_{xy}^{2\omega,\text{th}+\text{DL}} - a/(B + B_{\text{dem}} - B_{\text{ani}})$, as shown in Fig. 2f. Finally, by applying the geometry law, we can obtain the values of $R_{xx}^{2\omega,\text{th}} = L/W \cdot R_{xy}^{2\omega,\text{th}}$ (where *L* and *W* are the channel length and width of the fabricated Hall bar device) and $R_{xx}^{2\omega,MD-UMR}$ from Eq. (1). Consequently, Fig. 2g summarizes the current-dependent $R_{xx}^{2\omega}$, $R_{xx}^{2\omega,\text{th}}$, and , $R_{xx}^{2\omega,MD-UMR}$ curves, and the linear $R_{xx}^{2\omega,MD-UMR}$-*I* correlation validates the presence of MD-UMR effect in our CoFeB/InSb/CdTe system.

Moreover, to compare the MD-UMR effect in various systems, we adopted the normalized $R_{\text{normal}}^{MD-UMR} = R_{xx}^{2\omega,MD-UMR}/|R_{xx} \cdot \boldsymbol{j}|$ as the benchmark since it directly reflects the intrinsic MD-UMR strength. In this context, the amplitude of $R_{\text{normal}}^{MD-UMR}$ in our device is calibrated as $1.20 \times 10^{-11} A^{-1} cm^2$, which is 3-5 times larger than that in traditional FM/NM systems (*e.g.*, Co/Pt, Co/Ta, Co/W) at room temperature probably due to the larger SOC of InSb (Fig. 2h)[27]. Besides, we need to point out that although MD-UMR effects have also been discovered in topological insulators-based magnetic heterostructures with high $R_{\text{normal}}^{MD-UMR}$ values, yet such feature can only sustain at cryogenic temperatures (GaMnAs/$Bi_2Se_3$ at 30 K[28], CoFeB/$Bi_2Se_3$ at 150 K[14]), and the Cr-(Bi,Sb)$_2$Te$_3$/ (Bi,Sb)$_2$Te$_3$ system requires extra magnetic fields to align the original perpendicular magnetic anisotropy towards the in-plane direction[29, 30]. As a summary, the pronounced MD-UMR response observed in our CoFeB/InSb/CdTe heterostructures at room temperature paves way for constructing energy-efficient SOT devices.

In addition to the MD-UMR effect in the low **B**-field region, our previous study has also revealed an evident FD-UMR response in the InSb/CdTe heterostructures at room temperature[31]. As illustrated in Fig. 3a, the giant Rashba effect from the InSb/CdTe hetero-interface gives rise to a pseudomagnetic field $\bm{B}_{\text{eff}} = \alpha_R \cdot (\bm{k}_x \times \hat{\bm{z}})$, where $\alpha_R$ is the Rashba coefficient and $\bm{k}_x$ is the momentum operator of the conducting charge current. Based on the nonlinear second-order response model in band splitting material system[11], when the charge current conducts along the +*x*-direction at the InSb/CdTe interface, the Δ*k*-induced spin sub-band displacement (*i.e.*, it is noted that InSb has a negative *g*-factor) and the spin-momentum locking mechanism would polarize the electron spin toward +*y*-axis. Under such circumstances, the external in-plane magnetic field would break the time reversal symmetry and interact with the $\bm{B}_{\text{eff}}$-induced spin (+$\sigma_y$). As a result, the second-order charge current (*i.e.*, $I_c^{2\omega} \propto E_x^2 \sigma \cdot \bm{B}$, where $E_x$ is the electric field between the source/drain contacts) modifies the 2$^{\text{nd}}$-harmonic component of the InSb/CdTe channel resistance, and the resulting $R_{xx}^{2\omega,FD-UMR} = -2A \cdot R_{xx}^{\text{InSb}} \cdot \bm{B}_{\text{eff}} \cdot \bm{B}$ (*i.e.*, where *A* is a positive AMR coefficient of InSb[20]) reaches its highest (lowest) level when $\phi = 90°$ (270°)[20]. To elucidate the FD-UMR effect in our CoFeB/InSb/CdTe system, we subsequently expanded the applied magnetic field range as $\bm{B} = +B_y \cdot \hat{\bm{y}}$ where $-1\,\text{T} \leq B_y \leq +1\,\text{T}$. Strikingly, two distinctive features are captured from the measured field-dependent $R_{xx}^{2\omega}$ data shown in Figs. 3b-c. First of all, unlike conventional FM/NM counterparts whose second-order magnetoresistances remain constant in the high **B**-field region (*i.e.*, indicative of negligible FD-UMR contribution)[10, 32], the overall $R_{xx}^{2\omega}(+B_y, +I_x)$ curves of the CoFeB(5 nm)/InSb(15 nm)/CdTe Hall-bar device exhibit linear correlations with both the applied magnetic field and current, shown in Fig.3c (*i.e.*, we have also provided the magnetic field-dependent $R_{xx}^{2\omega}$ result of the InSb(15 nm)/CdTe control sample in Fig. 3b, and the same positive linear $R_{xx}^{2\omega} - B$ correlation again attests the same FD-UMR signature). Secondly, the magnetic field-induced UMR slope has an opposite polarity as compared with the MD-UMR response (Fig. 2c). In other word, the $R_{xx}^{2\omega}$ value gradually decreases towards the low-resistance state until the CoFeB layer is fully magnetized at $B_y$ = +2 mT (*i.e.*, magnetic saturation field), whereas it increases linearly in the high ***B***-field region. Moreover, the competition between the MD-UMR and FD-UMR components are also manifested from the evolution of the angular-dependent 2nd-harmonic magneto-resistance under different in-plane magnetic field. As displayed in Fig. 3d, the three sinusoidal UMR curves have the same period of 360°, yet their corresponding $R_{xx}^{2\omega}$ peak positions (*i.e.*, phases) shift from $\phi$ = 270° ($B_y$ = +50 and +300 mT) to $\phi$ = 90° ($B_y$ = +1000 mT), indicating that the dominant mechanism has transited from the CoFeB/InSb-associated MD-UMR to the InSb/CdTe interfacial Rashba-induced FD-UMR.

Finally, we demonstrated the operating principle of the CoFeB/InSb/CdTe-based two-terminal SOT device. Considering that sinusoid signals are not allowed in neither the write nor read process for conventional memory operation, we performed the DC measurement to readout the magnetic moment-dependent UMR information. Specifically, once the magnetization of the CoFeB layer was set by the initial ($B_y$, $I_x$) condition, a pair of channel magneto-resistances ($R_{xx}^{+I}$ and $R_{xx}^{-I}$) were recorded under constant positive and negative read currents of $I_{\text{read}}$ = ±10 mA (top panel of Fig. 4a). Afterwards, by subtracting these two datasets (*i.e.*, $R_{xx}^{\text{DC}} = (R_{xx}^{+I} - R_{xx}^{-I})/2$), we were able to re-construct the two-state $R_{xx}^{\text{DC}} - B$ hysteresis loop with a clockwise polarity, as shown in the bottom panel of Fig. 4a. Concurrently, the readout $R_{xx}^{\text{DC}}$ amplitude also display a high linearity with respect to the DC current (Fig. 4b), which is agreement with the AC measurement results. Next, the write operation of our two-terminal SOT device was visualized by MOKE microscopy. As exemplified in Fig. 4c, with the presence of a small in-plane assisted field of $B_{y0}$ = +1 mT, a positive write current of $I_{write}$ = +15 mA with the pulse width of 50 ms manages to switch of the magnetic moment of the CoFeB layer from the initial $-y$-axis (represented by the blue color) to the $+y$-axis (indicated by the red color), and the switching polarity complies with the SOT scenario that $\boldsymbol{B}_{SO} \propto \boldsymbol{\sigma} \times \boldsymbol{M}$ (where $\boldsymbol{B}_{SO}$ is the effective spin-orbit field, and $\boldsymbol{\sigma} = +\sigma_y \cdot \hat{y}$ is the electron spin accumulated at the CoFeB/InSb interface)[33]. More importantly, given the deterministic switching mechanism of the type-y IMA system[34], even the CoFeB layer is partially magnetized, the applied ($I_{write}$ = −15 mA, $B_y$ = −1 mT) condition still achieves a 100% full magnetization switching of the entire Hall-bar channel towards the $-y$-direction (the bottom panel of Fig. 4c), which implies that the write operation of our two-terminal SOT device can be realized without invoking the initialization process.

In conclusion, we have designed the CoFeB/InSb/CdTe heterostructures to host both the MD-UMR and FD-UMR effects at room temperature. Benefiting from the tunable nonreciprocal charge transport characteristic with the one-fold symmetric UMR line-shape, our proposed system can be incorporated into advanced magnetic sensors to enhance the directional sensitivity. Besides, given that the deterministic SOT-driven magnetization switching in the type-y IMA system does not need external magnetic field [34], we can, in principle, realize the field-free two-terminal SOT-MRAM prototype device by further optimizing the film stack structure (*e.g.*, change the CoFeB layer thickness to optimize its anisotropy). Our work showcases the heterostructure engineering as a suitable approach to integrate and broaden the functionality of the spintronic applications.

**Figure Legends/Captions**

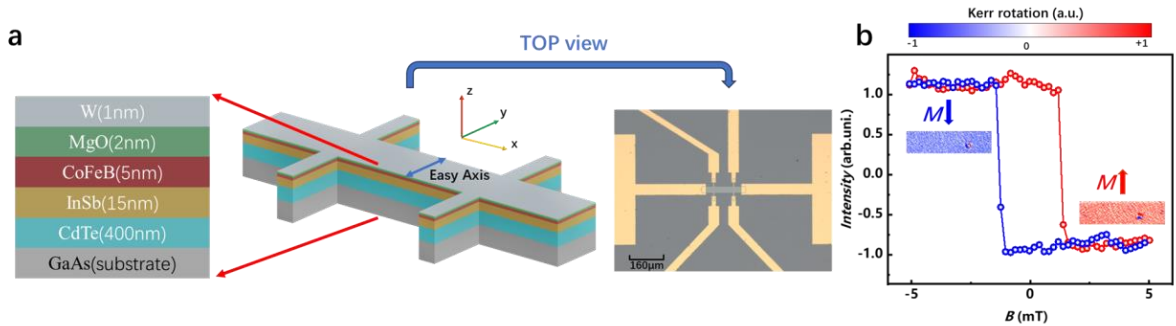

**Figure 1. Structural schematic and magnetic characterization of the CoFeB(5nm)/InSb (15 nm)/CdTe (400 nm) sample. a**. Schematic of the CoFeB/InSb/CdTe sample structure and the optical microscopy image of the fabricated six-probe Hall-bar device with the channel geometry of 30 μm (width) × 90 μm (length). Besides, the top W(1nm)/MgO(2nm) stack are used as the capping layer to protect sample, and the easy axis of CoFeB is along the *y*-axis (*i.e.*, perpendicular to the current direction). **b.** Magnetic field dependence of the light intensity with the MOKE configuration. Inset: Corresponding MOKE images of device in two fully magnetized regimes where the uniform red and blue patterns indicate the magnetization of +*y* (*$M_y$* ↑) and −*y* (*$M_y$* ↓) directions.

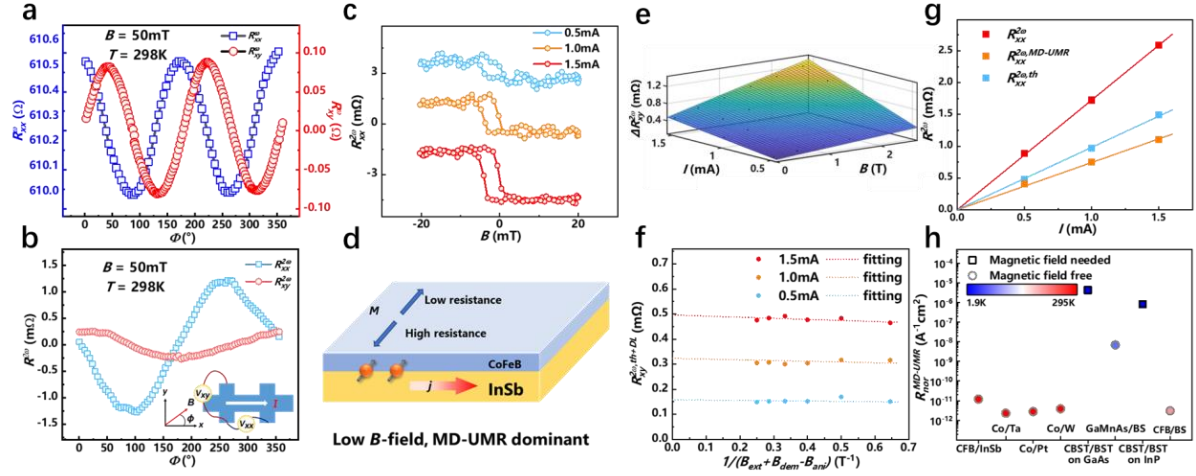

**Figure 2. Moment-dependent unidirectional magnetoresistance effect of the CoFeB(5nm)/InSb (15 nm)/CdTe (400 nm) sample in the low-field region at room temperature. a.** Angular-dependent first harmonic resistance ($R_{xx}^{\omega}$, $R_{xy}^{\omega}$) and **b.** second harmonic resistance ($R_{xy}^{2\omega}$, $R_{xy}^{2\omega}$) data with the presence of $|B|$ = 50 mT and $I_x$ = 1.5 mA. The inset image illustrates the measurement setup and the rotation angle $\phi$. **c.** Hysteresis $R_{xx}^{2\omega} - B_y$ curves under three current levels of $I_x$ = 0.5 mA, 1.0 mA and 1.5 mA, respectively. **d.** Illustration of the MD-UMR effect at the CoFeB/InSb interface. The parallel (anti-parallel) alignment between the spin polarization and the magnetic moment leads to the low (high) resistance state in reference to Fig. 2c. **e.** Three-dimensional mapping of the $\Delta R_{xy}^{2\omega}$ data under a set of external magnetic fields ($B$ = 0.05, 0.5, 1, 1.5, 2, and 2.5 T) and applied currents ($I$ = 0.5, 1.0, and 1.5 mA). **f.** Extraction of the thermal and damping-like contributions of $R_{xy}^{2\omega,th+DL}$ versus $1/(B + B_{\text{dem}} - B_{\text{ani}})$. The intercepts of the fitted dash lines represent the thermal components. **g.** Current dependence of $R_{xx}^{2\omega}$, $R_{xx}^{2\omega,th}$, and $R_{xx}^{2\omega,MD-UMR}$ of the CoFeB/InSb/CdTe sample. **h.** Comparison of the MD-UMR strength in various material systems. Circular (square) data points indicate that the MD-UMR effect is realized without (with) the presence of an external in-plane magnetic field. Our CoFeB/InSb/CdTe heterostructure stands out from the category by exhibiting a large field-free MD-UMR coefficient $R_{\text{normal}}^{\text{MD-UMR}}$ at room temperature.

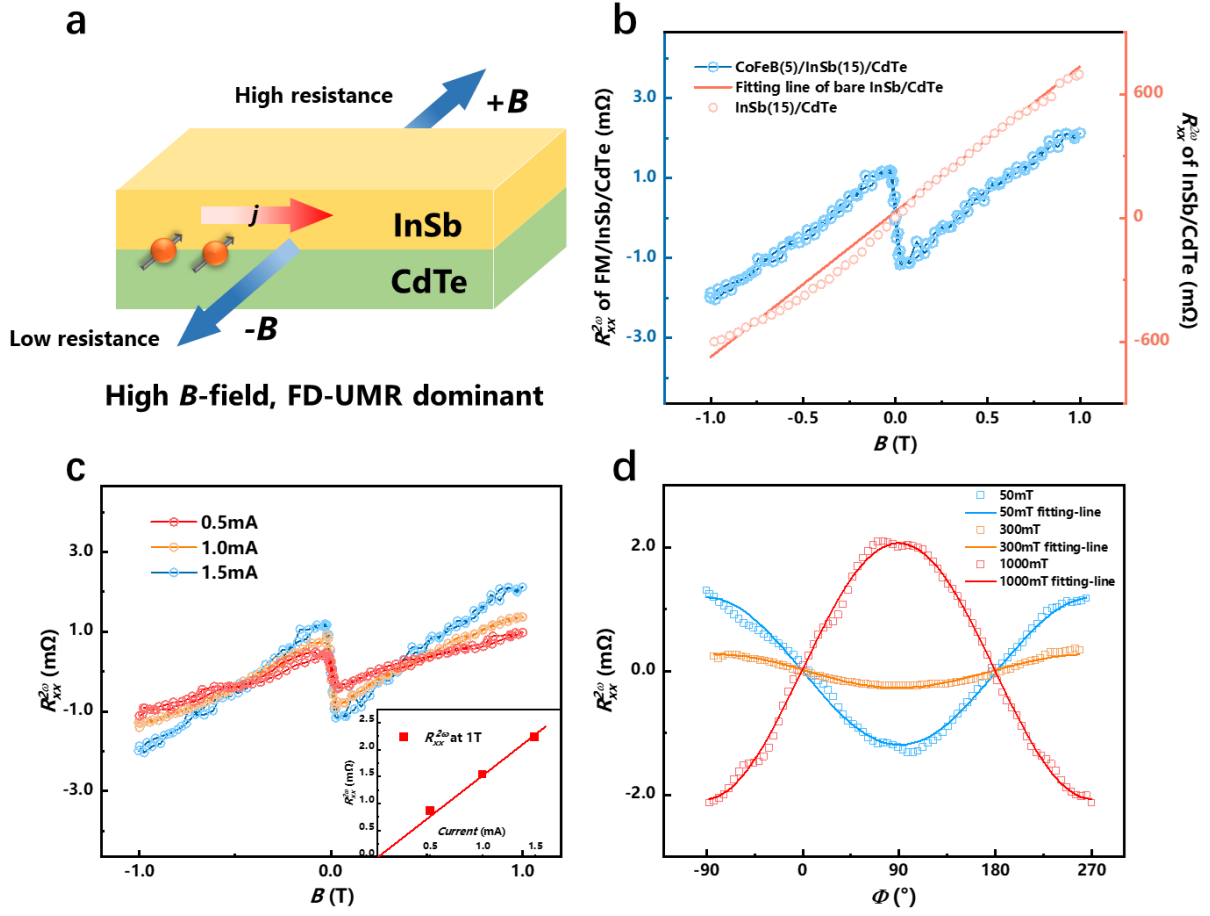

**Figure 3. Interplay between MD-UMR and FD-UMR effects in CoFeB/InSb/CdTe heterostructures. a.** Illustration of the field-driven unidirectional magnetoresistance effect at the InSb/CdTe interface. The high (low) resistance state corresponds to the parallel (anti-parallel) alignment between the polarized spin and the in-plane magnetic field. **b.** The measured $R_{xx}^{2\omega}$ data in the $B_y \subseteq [-1\ \text{T}, +1\ \text{T}]$ region. The CoFeB(15 nm)/CdTe(400 nm) control sample (red) also displays a linear $R_{xx}^{2\omega} - B$ relationship with the sample polarity. **c.** Room-temperature FD-UMR response under different current levels of 0.5 mA, 1.0 mA and 1.5 mA. **d.** Angular-dependent second-harmonic magnetoresistance $R_{xx}^{2\omega}$ at different magnetic fields of 50 mT, 300 mT, 1000 mT).

The overall UMR curve changes its peak/valley position at |B|>300 mT, implying the MD-UMR to FD-UMR transition in our CoFeB/InSb/CdTe system.

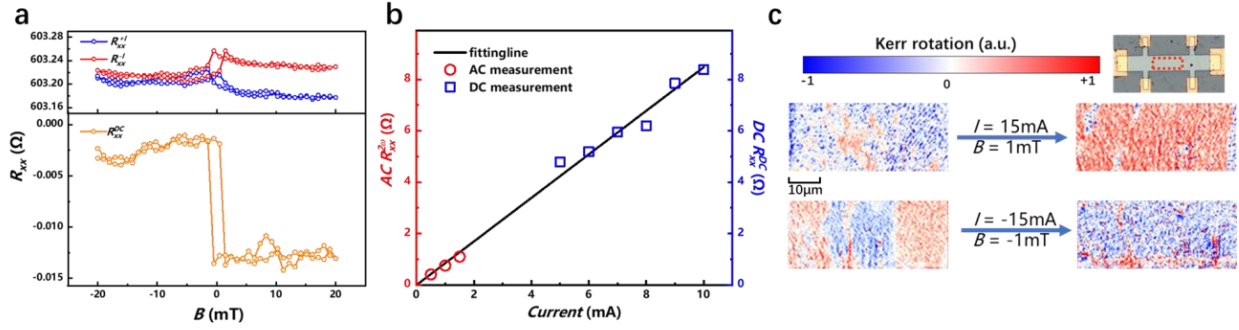

**Figure 4. Demonstration of the two-terminal read and write operations of the CoFeB/InSb/CdTe-based SOT device. a.** Magnetic-field dependent longitudinal resistance ($R_{xx}^{+I}$, $R_{xx}^{-I}$) recorded at $I_{DC} = \pm 10$ mA (top panel) and the corresponding two-state MD-UMR loop obtained by subtracting $R_{xx}^{DC} = (R_{xx}^{+I} - R_{xx}^{-I})/2$ (bottom panel) from the DC measurement. **b.** The MD-UMR signals from both the second harmonic ($R_{xx}^{2\omega}$) and DC ($R_{xx}^{DC}$) measurements follow the same linear $R_{xx}$–$I$ slope. **c.** Room-temperature SOT-driven magnetization switching of the CoFeB/InSb/CdTe device. The MOKE images in the bottom panel visualize the fully magnetized regimes after each write operation. A small in-plane magnetic field ($B_y = \pm 1$ mT) is introduced to facilitate the magnetization switching in the type-y IMA system as the CoFeB(5nm) is thick.

**References**

(1) Baibich, M. N.; Broto, J. M.; Fert, A.; Nguyen Van Dau, F.; Petroff, F.; Etienne, P.; Creuzet, G.; Friederich, A.; Chazelas, J. Giant magnetoresistance of (001)Fe/(001)Cr magnetic superlattices. *Phys Rev Lett* **1988**, *61* (21), 2472-2475. DOI: 10.1103/PhysRevLett.61.2472 From NLM Publisher.
(2) Binasch, G.; Grunberg, P.; Saurenbach, F.; Zinn, W. Enhanced magnetoresistance in layered magnetic structures with antiferromagnetic interlayer exchange. *Phys Rev B Condens Matter* **1989**, *39* (7), 4828-4830. DOI: 10.1103/physrevb.39.4828 From NLM PubMed-not-MEDLINE.
(3) Julliere, M. Tunneling between ferromagnetic films. *Physics letters A* **1975**, *54* (3), 225-226.
(4) Miyazaki, T.; Tezuka, N. Giant magnetic tunneling effect in Fe/Al2O3/Fe junction. *Journal of magnetism and magnetic materials* **1995**, *139* (3), L231-L234.
(5) Moodera, J. S.; Kinder, L. R.; Wong, T. M.; Meservey, R. Large magnetoresistance at room temperature in ferromagnetic thin film tunnel junctions. *Phys Rev Lett* **1995**, *74* (16), 3273-3276. DOI: 10.1103/PhysRevLett.74.3273 From NLM Publisher.
(6) Campbell, I.; Fert, A.; Jaoul, O. The spontaneous resistivity anisotropy in Ni-based alloys. *Journal of Physics C: Solid State Physics* **1970**, *3* (1S), S95.
(7) McGuire, T.; Potter, R. Anisotropic magnetoresistance in ferromagnetic 3d alloys. *IEEE Transactions on Magnetics* **1975**, *11* (4), 1018-1038.
(8) Miller, M.; Prinz, G.; Cheng, S.-F.; Bounnak, S. Detection of a micron-sized magnetic sphere using a ring-shaped anisotropic magnetoresistance-based sensor: A model for a magnetoresistance-based biosensor. *Applied Physics Letters* **2002**, *81* (12), 2211-2213.
(9) Jogschies, L.; Klaas, D.; Kruppe, R.; Rittinger, J.; Taptimthong, P.; Wienecke, A.; Rissing, L.; Wurz, M. C. Recent developments of magnetoresistive sensors for industrial applications. *Sensors* **2015**, *15* (11), 28665-28689.
(10) Avci, C. O.; Garello, K.; Ghosh, A.; Gabureac, M.; Alvarado, S. F.; Gambardella, P. Unidirectional spin Hall magnetoresistance in ferromagnet/normal metal bilayers. *Nature Physics* **2015**, *11* (7), 570-575. DOI: 10.1038/nphys3356.
(11) He, P.; Zhang, S. S. L.; Zhu, D.; Liu, Y.; Wang, Y.; Yu, J.; Vignale, G.; Yang, H. Bilinear magnetoelectric resistance as a probe of three-dimensional spin texture in topological surface states. *Nature Physics* **2018**, *14* (5), 495-499. DOI: 10.1038/s41567-017-0039-y.
(12) Tokura, Y.; Nagaosa, N. Nonreciprocal responses from non-centrosymmetric quantum materials. *Nature communications* **2018**, *9* (1), 3740.
(13) Olejník, K.; Novák, V.; Wunderlich, J.; Jungwirth, T. Electrical detection of magnetization reversal without auxiliary magnets. *Physical Review B* **2015**, *91* (18). DOI: 10.1103/PhysRevB.91.180402.
(14) Lv, Y.; Kally, J.; Zhang, D.; Lee, J. S.; Jamali, M.; Samarth, N.; Wang, J. P. Unidirectional spin-Hall and Rashba−Edelstein magnetoresistance in topological insulator-ferromagnet layer heterostructures. *Nat Commun* **2018**, *9* (1), 111. DOI: 10.1038/s41467-017-02491-3 From NLM PubMed-not-MEDLINE.
(15) Lou, K.; Zhao, Q.; Jiang, B.; Bi, C. Large Anomalous Unidirectional Magnetoresistance in a Single Ferromagnetic Layer. *Physical Review Applied* **2022**, *17* (6). DOI: 10.1103/PhysRevApplied.17.064052.
(16) Mehraeen, M.; Zhang, S. S. L. Spin anomalous-Hall unidirectional magnetoresistance. *Physical Review B* **2022**, *105* (18). DOI: 10.1103/PhysRevB.105.184423.

(17) Avci, C. O.; Mann, M.; Tan, A. J.; Gambardella, P.; Beach, G. S. A multi-state memory device based on the unidirectional spin Hall magnetoresistance. *Applied Physics Letters* **2017**, *110* (20).
(18) Liu, Y.-T.; Chen, T.-Y.; Lo, T.-H.; Tsai, T.-Y.; Yang, S.-Y.; Chang, Y.-J.; Wei, J.-H.; Pai, C.-F. Determination of spin-orbit-torque efficiencies in heterostructures with in-plane magnetic anisotropy. *Physical Review Applied* **2020**, *13* (4), 044032.
(19) Li, Y.; Li, Y.; Li, P.; Fang, B.; Yang, X.; Wen, Y.; Zheng, D.-x.; Zhang, C.-h.; He, X.; Manchon, A. Nonreciprocal charge transport up to room temperature in bulk Rashba semiconductor α-GeTe. *Nature communications* **2021**, *12* (1), 540.
(20) Guillet, T.; Zucchetti, C.; Barbedienne, Q.; Marty, A.; Isella, G.; Cagnon, L.; Vergnaud, C.; Jaffres, H.; Reyren, N.; George, J. M.; et al. Observation of Large Unidirectional Rashba Magnetoresistance in Ge(111). *Phys Rev Lett* **2020**, *124* (2), 027201. DOI: 10.1103/PhysRevLett.124.027201 From NLM PubMed-not-MEDLINE.
(21) Gmitra, M.; Fabian, J. First-principles studies of orbital and spin-orbit properties of GaAs, GaSb, InAs, and InSb zinc-blende and wurtzite semiconductors. *Physical Review B* **2016**, *94* (16), 165202.
(22) Li, J.; Tang, C.; Du, P.; Jiang, Y.; Zhang, Y.; Zhao, X.; Gong, Q.; Kou, X. Epitaxial growth of lattice-matched InSb/CdTe heterostructures on the GaAs (111) substrate by molecular beam epitaxy. *Applied Physics Letters* **2020**, *116* (12).
(23) Zhang, Y.; Xue, F.; Tang, C.; Li, J.; Liao, L.; Li, L.; Liu, X.; Yang, Y.; Song, C.; Kou, X. Highly efficient electric-field control of giant Rashba spin–orbit coupling in lattice-matched InSb/CdTe heterostructures. *ACS nano* **2020**, *14* (12), 17396-17404.
(24) Avci, C. O.; Garello, K.; Mendil, J.; Ghosh, A.; Blasakis, N.; Gabureac, M.; Trassin, M.; Fiebig, M.; Gambardella, P. Magnetoresistance of heavy and light metal/ferromagnet bilayers. *Applied Physics Letters* **2015**, *107* (19).
(25) Yin, Y.; Han, D.-S.; de Jong, M. C.; Lavrijsen, R.; Duine, R. A.; Swagten, H. J.; Koopmans, B. Thickness dependence of unidirectional spin-Hall magnetoresistance in metallic bilayers. *Applied Physics Letters* **2017**, *111* (23).
(26) Zhou, X.; Zeng, F.; Jia, M.; Chen, H.; Wu, Y. Sign reversal of unidirectional magnetoresistance in monocrystalline Fe/Pt bilayers. *Physical Review B* **2021**, *104* (18), 184413.
(27) Kane, E. O. Band structure of indium antimonide. *Journal of Physics and Chemistry of Solids* **1957**, *1* (4), 249-261.
(28) Duy Khang, N. H.; Hai, P. N. Giant unidirectional spin Hall magnetoresistance in topological insulator – ferromagnetic semiconductor heterostructures. *Journal of Applied Physics* **2019**, *126* (23). DOI: 10.1063/1.5134728.
(29) Yasuda, K.; Tsukazaki, A.; Yoshimi, R.; Takahashi, K. S.; Kawasaki, M.; Tokura, Y. Large Unidirectional Magnetoresistance in a Magnetic Topological Insulator. *Phys Rev Lett* **2016**, *117* (12), 127202. DOI: 10.1103/PhysRevLett.117.127202 From NLM PubMed-not-MEDLINE.
(30) Fan, Y.; Shao, Q.; Pan, L.; Che, X.; He, Q.; Yin, G.; Zheng, C.; Yu, G.; Nie, T.; Masir, M. R.; et al. Unidirectional Magneto-Resistance in Modulation-Doped Magnetic Topological Insulators. *Nano Lett* **2019**, *19* (2), 692-698. DOI: 10.1021/acs.nanolett.8b03702 From NLM PubMed-not-MEDLINE.
(31) Li, L.; Wu, Y.; Liu, X.; Liu, J.; Ruan, H.; Zhi, Z.; Zhang, Y.; Huang, P.; Ji, Y.; Tang, C. Room-Temperature Gate-Tunable Nonreciprocal Charge Transport in Lattice-Matched InSb/CdTe Heterostructures. *Advanced Materials* **2023**, *35* (3), 2207322.

(32) Avci, C. O.; Mendil, J.; Beach, G. S. D.; Gambardella, P. Origins of the Unidirectional Spin Hall Magnetoresistance in Metallic Bilayers. *Phys Rev Lett* **2018**, *121* (8), 087207. DOI: 10.1103/PhysRevLett.121.087207 From NLM PubMed-not-MEDLINE.

(33) Sinova, J.; Valenzuela, S. O.; Wunderlich, J.; Back, C.; Jungwirth, T. Spin hall effects. *Reviews of modern physics* **2015**, *87* (4), 1213.

(34) Fukami, S.; Anekawa, T.; Zhang, C.; Ohno, H. A spin–orbit torque switching scheme with collinear magnetic easy axis and current configuration. *nature nanotechnology* **2016**, *11* (7), 621-625.